\documentclass[11pt,nofootinbib,aps,pre]{revtex4-1}
\usepackage{amsmath,amsfonts,amssymb}
\usepackage{graphicx}
\usepackage{algorithmic}

\newcommand{\arctanh}[1]{\mathrm{arctanh}#1}
\newcommand{\Sign}[1]{\mathrm{Sign}#1}

\newcommand{\Jij}{J_{ij}}
\newcommand{\ie}{\textit{i.e.\ }}

\newcommand{\Ham}{\mathcal{H}}
\newcommand{\B}{\beta}
\newcommand{\rref}[1]{Eq.~(\ref{#1})}
\newcommand{\<}{\langle}
\renewcommand{\>}{\rangle}
\newcommand{\cP}{\mathcal{P}}
\newcommand{\cL}{\mathcal{L}}
\newcommand{\cR}{\mathcal{R}}
\newcommand{\cU}{\mathcal{U}}
\newcommand{\cD}{\mathcal{D}}

\begin{document}

\title{A very fast inference algorithm for finite-dimensional spin
  glasses: Belief Propagation on the dual lattice.}

\author{Alejandro Lage-Castellanos, Roberto Mulet}
\affiliation{Department of Theoretical Physics, Physics Faculty,
  University of Havana, La Habana, CP 10400, Cuba. }

\author{Federico Ricci-Tersenghi} \affiliation{Dipartimento di Fisica,
  INFN -- Sezione di Roma 1 and CNR -- IPCF, UOS di Roma,\\
  Universit\`{a} La Sapienza, P.le A. Moro 5, 00185 Roma, Italy}

\author{Tommaso Rizzo} \affiliation{Dipartimento di Fisica,
  Universit\`{a} La Sapienza, P.le A. Moro 5, 00185 Roma, Italy}

\date{\today}

\begin{abstract}
  Starting from a Cluster Variational Method, and inspired by the
  correctness of the paramagnetic Ansatz (at high temperatures in
  general, and at any temperature in the 2D Edwards-Anderson model) we
  propose a novel message passing algorithm --- the Dual algorithm ---
  to estimate the marginal probabilities of spin glasses on finite
  dimensional lattices. We show that in a wide range of temperatures
  our algorithm compares very well with Monte Carlo simulations, with
  the Double Loop algorithm and with exact calculation of the ground
  state of 2D systems with bimodal and Gaussian interactions.
  Moreover it is usually 100 times faster than other provably
  convergent methods, as the Double Loop algorithm.
\end{abstract}

\maketitle

\section{Introduction}

Inference problems are common to almost any scientific discipline.
Often an inference problem can be recast in that of computing some
marginal probability on a small subset of variables, given a joint
probability distribution over a large number $N$ of variables,
$\{s_1,\ldots,s_N\} \equiv \sigma$. A typical example is the
computation of the expectation value of a variable $\langle s_i
\rangle \equiv \sum_{s_i} s_i\, p_i(s_i)$ where $p_i(s_i)$ is the
single-variable marginal probability defined as
\[
p_i(s_i) \equiv \sum_{\sigma \setminus s_i} P(\sigma)\;.
\]
Clearly the computation of the sum in the r.h.s.\ is as difficult as
the computation of the partition function $Z \equiv \sum_\sigma
P(\sigma)$, which is the main subject of statistical mechanics.  In
the general (and the most interesting) case, this problem can not be
solved exactly in a time growing sub-exponentially with the size $N$.
We are deemed to use some kind of approximation in order to compute
marginals in a time growing linearly in $N$.  The approximation
schemes used so far are mainly adopted from the field of statistical
mechanics \cite{libroMarcAndrea}, where mean-field-like approximations
are standard and well controlled tools for approximating the
free-energy, $-\beta f = \ln Z$.

For non-disordered models, like the Ising ferromagnetic model, these
approximations work quite well and provide good results
\cite{pelizzola}.  However, much less is known for models with
disorder, and for this reason we will focus on spin glasses in the
present paper.  In disordered models one usually deals with an
ensemble of problems (e.g.\ in spin glasses each sample has its own
couplings, randomly chosen from a given distribution), and the results
obtained by the statistical mechanics tools refer to average
quantities, i.e. those of the typical samples.  In other words one is
not concerned with the behavior of a specific sample, but rather one
looks at the whole ensemble.  On the contrary when doing inference one
is interested on the properties of a single specific problem and thus
the above approximation schemes (based on statistical mechanics
mean-field approaches) need to be converted into an algorithm that can
be run on such a specific sample.  Computing marginals on a given
sample clearly gives more information than computing averages over the
ensemble.

To our knowledge, effective (i.e. linear time in $N$) algorithms for
computing marginals can be essentially divided in two broad classes:
stochastic local search algorithms, that roughly sample the
configurational space according to $P(\sigma)$, and algorithms based
on some kind of mean-field approximation.  The former are exact on the
long run, but the latter can be much more useful if an approximated
answer is required in a short time.  Unfortunately the latter also
have some additional drawback due to the mean-field nature of the
underlying approximation, e.g. the appearance of spurious phase
transitions, that may prevent the proper convergence of the algorithm.

One more reason why this latter class of algorithms has a limited
scope of application is that the convergence of the algorithm may
strongly depend on the presence of short loops in the network of
variables interactions.  In this sense the successful application of
one of these algorithms to models defined on regular lattices (which
have many short loops) would be a major achievement.

In this paper we introduce a fast algorithm for computing marginals in
2D and 3D spin glass models \cite{libroYoung} defined by the
Hamiltonian (further details are given below)
\begin{equation}
\Ham (\sigma) = - \sum_{(i,j)} \Jij s_i s_j\;.
\label{Ham}
\end{equation}

The first non trivial mean-field approximation for the above model
corresponds to the Bethe-Peierls approximation scheme and the Belief
Propagation (BP) algorithm \cite{kabaSaad}. Unfortunately when BP is
run on a given spin glass sample defined on a $D$ dimensional lattice,
it seems to provide exactly the same output as if it were run on a
random regular graph with fixed degree $2D$: that is, for $T \gtrsim
T_\text{Bethe}$ it converges to a solution with all null local
marginals ($\langle s_i \rangle = 0$), while for $T \lesssim
T_\text{Bethe}$ it does not converge to a fixed point.

The next step in the series of mean-field approximations (also known
as Kikuchi approximations or `cluster variation method') is to
consider joint probability distributions of the four spins belonging
to the same plaquette \cite{kikuchi,pelizzola}. Under this
approximation an algorithm has been derived which is called
Generalized Belief Propagation (GBP) \cite{yedidia}.  To our
knowledge, this algorithm has been applied to 2D spin glasses, but
only in presence of an external magnetic field, which is known to
improve the convergence properties of GBP.  Our experience says that,
running plain GBP on a generic sample of a 2D spin glass without
external field, a fixed point is reached only for high enough
temperatures, well above the interesting region.  The lack of
convergence of GBP (and other similar message-passing algorithms, MPA)
is a well known problem, whose solution is in general far from being
understood.  For this reason, a new class of algorithms has been
recently introduced \cite{HAK03}, which provably converge to a fixed
point: these algorithms use a double loop iterative procedure (to be
compared to the single loop in GBP) and for this reason they are
usually quite slow.

The algorithm we are going to introduce is as fast as BP (which is the
fastest algorithm presenty available), and converges in a wider range
of temperatures than BP. Moreover the marginal probabilities provided
by our algorithm are as accurate as those which can be obtained by
Double Loop algorithms in a much larger running time.

Our algorithm works in the absence of external field, that is in the
situation where MPA have more problems in converging to a fixed point.
In this sense it is a very important extension to presently available
inference algorithms.

The rest of the work is organized as follows.  First, in section
\ref{GBP2D} we derive the GBP equations for the 2D Edward Anderson
model. Then, in section \ref{sec:multfields} we rewrite these
equations in terms of fields, a notation that has a nicer physical
interpretation and that we are going to use in the rest of the
work. Section \ref{Dual} presents the novel algorithm, inspired by the
paramagnetic Ansatz to the GBP equations.  In section \ref{Results2D}
we show the results of running this algorithm on the 2D
Edward-Anderson model. There we compare its performance with Monte
Carlo simulations, with the Double Loop algorithm and with exact
calculation of the ground state of systems with bimodal and Gaussian
interactions. Then, in section \ref{Results3D} we generalized our
message passing equation to general dimensions and present some
results for the 3D Edward-Anderson model. Finally, some conclusions
are drawn in section \ref{Conclusions}.

\section{Generalized Belief Propagation on the 2D EA model}
\label{GBP2D}

Here we present the GBP equations for
the Edwards-Anderson (EA) model on a 2D square lattice, and we refer
the reader to \cite{yedidia,pelizzola} for a more general
introduction.  In our case (as well as in many other cases) GBP is
equivalent to Kikuchi's approximation, known as the Cluster
Variational Method (CVM) \cite{kikuchi}. We will try a presentation as
physical as possible.

Consider the 2D EA model consisting of a set $\sigma=\{s_1, \ldots,
s_N\}$ of $N$ Ising spins $s_i=\pm 1$ located at the nodes of a 2D
square lattice, interacting with a Hamiltonian
\begin{equation}
\Ham (\sigma) = - \sum_{\langle i,j \rangle} \Jij s_i s_j\;,
\label{eq:Ham}
\end{equation}
where the sum runs over all couples of neighboring spins (first
neighbors on the lattice). The $\Jij$ are the coupling constants
between spins and are supposed to be fixed for any given instance of
the model. If the interactions are not random variables, \ie $\Jij=J$,
then the 2D ferromagnet is recovered.  We will focus on the two most
common disorder distributions: bimodal interactions, $P(J) =
\frac{1}{2} \delta(J-1)+\frac{1}{2} \delta(J+1)$, and Gaussian
interactions $P(J) = \exp(-J^2/2)/\sqrt{2 \pi}$.
 
The statistical mechanics of the EA model, at a given temperature
$T=1/\B $, is given by the Gibbs-Boltzmann distribution
\[
P(\sigma) = \frac{e^{-\B\Ham(\sigma)}}{Z}\;.
\]
The direct computation of the partition function
\[
Z=\sum_{\sigma} e^{-\B \Ham(\sigma)}
\]
or any marginal distribution $p(s_i,s_j)=\sum_{\sigma \setminus
  s_i,s_j}P(\sigma)$ is a time consuming task, unattainable in
practice, since it involves the addition of $2^N$ terms, and therefore
an approximation is required\footnote{The 2D case is special: indeed,
  thanks to the small genus topology, the partition function $Z$ can
  be computed efficiently. However we are interested in developing an
  algorithm for the general case, and we will not make use of this
  peculiarity.}.

The idea of the Region Graph Approximation to Free Energy
\cite{yedidia} is to replace the real distribution $P(\sigma)$ by a
reduced set of its marginals. The hierarchy of approximations is given
by the size of such marginals, starting with the set of all single
spins marginals $p_i(s_i)$ [mean-field], then following to all
neighboring sites marginals $p_{ij}(s_i,s_j)$ [Bethe], then to all
square plaquettes marginals $p_{ijkl}(s_i,s_j,s_k,s_l)$, and so
on. Since the only way of knowing such marginals exactly is the
unattainable computation of $Z$, the method pretends to approximate
them by a set of beliefs $b_i(s_i)$, $b_{ij}(s_i,s_j)$, etc.\ obtained
from the minimization of a region based free energy. In the Region
Graph Approximation to Free Energy, a set of regions, \ie sets of
variables and their interactions, is defined, and a free energy is
written in terms of the beliefs at each region. The Cluster
Variational Method does a similar job, but instead of starting from an
arbitrary choice of regions, it starts by defining the set of largest
regions, and smaller regions are defined recursively by the
intersections of bigger regions. In this sense, CVM is a specific
choice of region graph approximation to the free energy.

For the 2D EA model, we will consider the expansion of the free energy
in terms of the marginals at three levels of regions: single sites (or
spins), links, and plaquettes. By plaquettes we mean the square basic
cell of the 2D lattice. This choice of regions corresponds to the CVM
having the square plaquettes as biggest regions. The free energy of
the system is therefore written as
\[
F=\sum_{R}c_R  F_R\;,
\]
where $R$ runs over all regions considered, and the free energy in a
particular region depends on the marginals at that level
$b_R(\sigma_R)$:
\[
\B F_R = \sum_{\sigma_R} b_R(\sigma_R) \B E_R(\sigma_R) +
b_R(\sigma_R) \log b_R(\sigma_R)\;.
\]
The symbol $\sigma_R$ refers to the set of spins in region $R$, while
$E_R$ is the energy contribution in that region. The counting numbers
$c_R$ (also Moebius coefficients) are needed to ensure that bigger
regions do not over count the contribution in free energy of smaller
regions, and follow the prescription
\begin{equation}
c_R = 1- \sum_{R'\supset R} c_{R'}\;,
\label{eq:countingnumbers}
\end{equation}
where $R'$ is any region containing completely region $R$, as e.g.\ a
plaquette containing a link or a link containing a site.  In the case
of the 2D lattice, the biggest regions are the square plaquettes, and
therefore $c_\text{plaq} = 1$, while the links regions have
$c_\text{link} = 1 - 2 c_\text{plaq} = -1$ (as each of them is
contained in 2 plaquettes regions), and finally the spins regions have
$c_\text{site} = 1 - 4 c_\text{plaq} - 4 c_\text{link} = 1$ (as each
spin belongs to 4 links and 4 plaquettes). So the actual approximation
for the EA model on a 2D square lattice is
\begin{eqnarray}
\B F &=& \sum_{\cP} \sum_{\sigma_\cP} b_\cP(\sigma_\cP)
  \log \frac{b_\cP(\sigma_\cP)}{\exp[-\B E_\cP(\sigma_\cP)]}
  \qquad \text{Plaquettes} \nonumber\\
& & -\sum_{L} \sum_{\sigma_L} b_L(\sigma_L)
  \log \frac{b_L(\sigma_L)}{\exp[-\B E_L(\sigma_L)]}
  \qquad \text{Links} \label{eq:freeen} \\
& & +\sum_{i} \sum_{s_i} b_i(s_i)
  \log \frac{b_i(s_i)}{\exp[-\B E_i(s_i)]}
  \qquad\text{Sites} \nonumber
\end{eqnarray}
where the sums run over all plaquettes, links and sites respectively.
Please note that we are using the following notation for region
indices: lower-case for sites, upper-case for links and upper-case
calligraphic for plaquettes.  The energy term $E_i(s_i)$ in the sites
contribution is only relevant when an external field acts over spins,
and can be taken as zero in our case, since no external field is
considered. Notice that whenever the interactions are included in more
than one region (in our case are included in Link and Plaquette
regions), the counting numbers guarantee that the exact
thermodynamical energy $U=\sum_\sigma P(\sigma) \Ham(\sigma)$ is
obtained when the beliefs are the exact marginals of the Boltzmann
distribution. On the other hand, the entropy contribution is
intrinsically approximated, since the cutoff in the regions sizes
imposes a certain kind of factorization of $P(\sigma)$ in terms of its
marginals (see \cite{pelizzola} for an explanation of the region graph
approximation in terms of cumulants expansions of the entropy).

The next step in the method, is to compute the beliefs from the
minimization condition of the free energy. However, an unrestricted
minimization will generally produce inconsistent solutions, since the
beliefs (marginals) are not independent, as they are related by the
marginalization conditions
\begin{equation}
\begin{array}{l}
b_i(s_i) = \sum_{\sigma_{L \setminus i}} b_L(\sigma_L) = \sum_{s_j}
b_L(s_i,s_j)\;,\\
b_L(\sigma_L) = b_L(s_i,s_j) = \sum_{\sigma_{\cP \setminus L}}
b_\cP(\sigma_\cP) = \sum_{s_k,s_l} b_\cP(s_i,s_j,s_k,s_l)\;,
\end{array}
\label{eq:marginalization}
\end{equation}
where $\sigma_L=\{s_i,s_j\}$ and $\sigma_\cP=\{s_i,s_j,s_k,s_l\}$.  In
order to minimize under the constraints in \rref{eq:marginalization}
and under the normalization condition for each belief, a set of
Lagrange multipliers should be added to the free energy in
\rref{eq:freeen}. There are different ways of choosing the Lagrange
multipliers \cite{yedidia}, and each of them will produce a different
set of self consistency equations. We choose the so called Parent to
Child scheme (see section XX in \cite{yedidia}), in which constraints
in \rref{eq:marginalization} are imposed by two sets of Lagrange
multipliers: $\mu_{L\to i}(s_i)$ relating the belief at link $L$ to
that at site $i$, and $\nu_{\cP\to L}(\sigma_L)$ relating the one at
plaquette $\cP$ to the one at link $L$.

\begin{figure}[htb]
\begin{center}              
\includegraphics[width=0.9\textwidth]{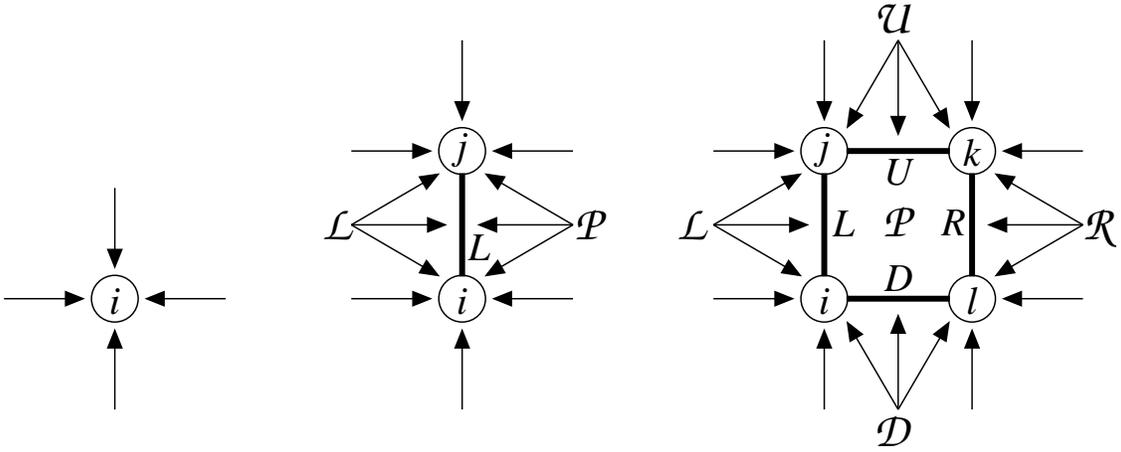}
\caption{Schematic representation of belief equations
  (\ref{eq:beliefs}).  Lagrange multipliers are depicted as arrows,
  going from parent regions to children regions.}
\label{fig:beliefs2D}
\end{center} 
\end{figure}

With constraints (\ref{eq:marginalization}) enforced by Lagrange
multipliers, the free energy stationary conditions for the beliefs are
the following:
\begin{eqnarray}
b_i(s_i) &=& \frac{1}{Z_i} \exp\left(-\B E_i(s_i) - \sum_{L\supset
  i}^4 \mu_{L\to i}(s_i)\right)\;, \nonumber\\
b_L(\sigma_L) &=& \frac{1}{Z_L} \exp\left(-\B E_L(\sigma_L) -
\sum_{\cP \supset L}^2 \nu_{\cP\to L}(\sigma_L) - \sum_{i\subset L}^2
\mathop{\sum_{L'\supset i}^3}_{L' \neq L} \mu_{L'\to i}(s_i)
\right)\;, \label{eq:beliefs}\\
b_\cP(\sigma_\cP) &=& \frac{1}{Z_\cP} \exp\left(-\B E_\cP(\sigma_\cP)
- \sum_{L\subset\cP}^4 \mathop{\sum_{\cP'\supset L}^1}_{\cP' \neq \cP}
\nu_{\cP'\to L}(\sigma_L) - \sum_{i\subset \cP}^4 \mathop{\sum_{L \supset
    i}^2}_{L \not\subset \cP} \mu_{L\to i}(s_i) \right)\;, \nonumber
\end{eqnarray}
where the notation $L\supset i$ refers to all links containing site
$i$ and $\cP\supset L$ to all plaquettes containing link $L$.  The
upper indices in the sums are written just to help understanding how
many terms are in each sum for the 2D case.  The precise meaning of
the indices in each summation can be understood from the graphical
representation in Figure \ref{fig:beliefs2D}.  Lagrange multipliers
are shown as arrows going from parent regions to children regions:
simple arrows correspond to $\mu_{L\to i}$ and triple arrows to
$\nu_{\cP\to L}$. Let us consider, for instance, the belief in a link
region $b_L(\sigma_L)$, depicted in the central picture of
Fig.~\ref{fig:beliefs2D}: the sum of the two Lagrange multipliers
$\nu_{\cP\to L} (\sigma_L)$ corresponds to the triple arrows from
plaquettes on the left and right of the link $L$, while the double sum
over the three $\mu_{L\to i}(s_i)$ and the three $\mu_{L\to j}(s_j)$
correspond to the three arrows acting over the two spins.

In Eq.(\ref{eq:beliefs}), the $Z_R$ are normalization constants, and
the terms $E_\cP(\sigma_\cP)=E_\cP(s_i,s_j,s_k,s_l)=-(J_{ij} s_i s_j +
J_{jk} s_j s_k + J_{kl} s_k s_k + J_{li} s_l s_i)$ and $E_L(\sigma_L)
= E_L(s_i,s_j)=- J_{ij} s_i s_j$ are the corresponding energies in
plaquettes and links, and are represented in Fig.~\ref{fig:beliefs2D}
by bold lines (interactions) between circles (spins). In our case
$E_i(s_i)$ is zero since no external field is acting on the spins.

\begin{figure}[htb]
\begin{center}              
\includegraphics[width=0.9\textwidth]{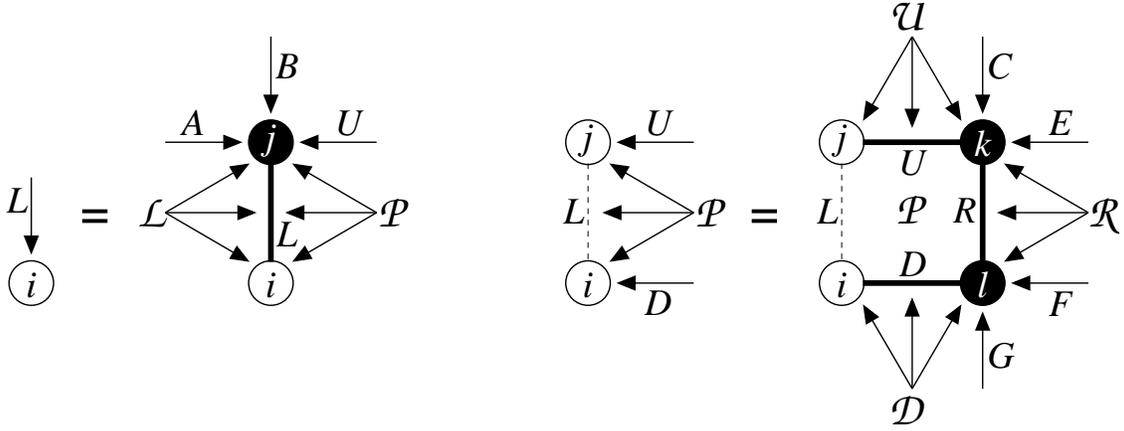} 
\caption{Message passing equations (\ref{eq:message2D}), shown
  schematically.  Messages are depicted as arrows, going from parent
  regions to children regions. On any link $\Jij$, represented as bold
  lines between spins (circles), a Boltzmann factor $e^{\B \Jij s_i
    s_j}$ exists. Dark circles represent spins to be traced
  over. Messages from plaquettes to links $\nu_{P\to L}(s_i,s_j)$ are
  represented by a triple arrow, because they can be written in terms
  of three parameters $U$, $u_i$ and $u_j$, defining the correlation
  $\< s_i s_j \>$ and magnetizations $\< s_i \>$ and $\< s_j \>$,
  respectively.}
\label{fig:message2D}
\end{center} 
\end{figure}

The Lagrange multipliers are fixed by the constraints they were
supposed to enforce, \rref{eq:marginalization}, and they must satisfy
the following set of self-consistency equations:
\begin{eqnarray}
& \exp\!\! & \Big[-\mu_{L\to i}(s_i)\Big] = \sum_{s_j} \exp\Bigg[-\B
    E_{L\setminus i}(s_i,s_j) - \sum_{\cP \supset L}^2 \nu_{\cP\to
      L}(s_i,s_j) - \mathop{\sum_{L'\supset j}^3}_{L' \neq L}
    \mu_{L'\to j}(s_j) \Bigg] \nonumber \\
& \exp\!\! & \Big[ -\nu_{\cP\to L}(s_i,s_j) - \mu_{D\to i}(s_i) -
    \mu_{U\to j}(s_j)\Big] = \label{eq:message2D} \\
&& \sum_{s_k,s_l} \exp \Bigg[-\B E_{\cP\setminus L}(s_i,s_j,s_k,s_l) -
    \mathop{\sum_{L' \in \cP}^3}_{L' \neq L} \mathop{\sum_{\cP'
        \supset L'}^1}_{\cP' \neq \cP} \nu_{P'\to L'}(\sigma_{L'}) -
    \mathop{\sum_{L' \supset k}^2}_{L'\not\subset\cP} \mu_{L'\to
      k}(s_k) - \mathop{\sum_{L' \supset l}^2}_{L'\not\subset\cP}
    \mu_{L'\to l}(s_l) \Bigg]\nonumber
\end{eqnarray}
Again, to help understanding these equations, we provide in
Fig.~\ref{fig:message2D} their graphical representation. Note that
there is one of these equations for every pair of Link-Site and every
pair of Plaquette-Link in the graph. With $E_{P\setminus L}$ we refer
to interactions in plaquette $P$ that are not in link $L$.

For each link $L$ in the 2D lattice, there are two link-to-site
multipliers, $\mu_{L\to i} (s_i)$ and $\mu_{L\to j} (s_j)$. For each
plaquette there are four plaquette-to-link multipliers $\nu_{P\to
  L}(s_i,s_j)$, corresponding to the four links contained inside the
plaquette. Let $N$ be the number of spins in the lattice, there are $2
N$ links and $N$ plaquettes. So the originally intractable problem of
computing marginals, has been replaced by the problem of solving a set
of $4 N + 4 N$ coupled equations for the Lagrange multipliers as those
in \rref{eq:message2D}.  Once these equations are solved, the
approximation for the marginals is obtained from \rref{eq:beliefs} for
the beliefs, and all thermodynamic quantities are derived from them as
in \rref{eq:freeen}.

Minimizing a region graph approximation to free energy, as that in
\rref{eq:freeen} with constraints \rref{eq:marginalization}, or
equivalently solving the set of self-consistent equations in
\rref{eq:message2D}, is still a non trivial task. Let us consider two
ways of doing it. The first method is the ``direct'' minimization of
the constrained free energy, using a Double Loop algorithm
\cite{HAK03}. This method is quite solid, since it guarantees
convergence to an extremal point of the constrained free energy, but
it may be very slow to converge. The second method, which is generally
faster but is not guaranteed to converge, is the family of the so
called Message-Passing algorithm (MPA), in which the Lagrange
multipliers are interpreted as \textit{messages} $\nu_{\cP\to
  L}(\sigma_L)$ going from plaquettes to links, and messages
$\mu_{L\to i}(s_i)$ from links to sites. Self consistency equations
(\ref{eq:message2D}) can be viewed as the update rules for the
messages in the left hand side, in terms of those in the right hand
side. A random order updating of the messages in the graph by
\rref{eq:message2D} (message passing) can reach a fixed point
solution, and therefore, to an extremal point of the constrained free
energy \cite{yedidia}. Next, we show explicitly how the
message-passing equations looks like in terms of fields.

\subsection{From multipliers to fields}
\label{sec:multfields}

A particularly useful way of representing the multipliers (messages),
with a nice physical interpretation, is the one used in
\cite{kabashima}, which we adopt here. In full generality
\cite{kabashima, TLMF10}, these multipliers can be written in terms of
effective fields
\begin{eqnarray}
\mu_{L\to i}(s_i) &=& \B\, u_{L\to i}\,s_i \\
\nu_{\cP\to L}(s_i,s_j) &=& \B\,(U_{\cP\to L}\,s_i\,s_j + u_{\cP\to
  i}\,s_i + u_{\cP\to j}\,s_j)
\end{eqnarray}
In particular, the field $u$ corresponds to the cavity field in the
Bethe approximation \cite{yedidia}. Using Lagrange multipliers,
messages or fields, is essentially equivalent. We will often refer to
fields as $u$-messages to emphasize their role in a message-passing
algorithm, and we will refer to self-consistency equations
(\ref{eq:message2D}) as the message-passing equations.

This parametrization of the multipliers has proved useful to other
endeavors, like the extension of the replica theory to general region
graph approximations \cite{TLMF10}. Here, all the relevant information
in the Lagrange multipliers is translated to ``effective fields'' $u$
and $(U,u_a, u_b)$. Notice that in this representation every single
field $u$ corresponds to an arrow in the schematic messages-passing
equations in Figure \ref{fig:message2D}. In particular, the messages
going from plaquettes to links are characterized by three fields
$(U,u_a,u_b)$, and the field $U$ acts as an effective interaction
term, that adds directly to the energy terms in the Boltzmann
factor. For instance, the first message-passing \rref{eq:message2D} is
\begin{align}
\exp \Big[ \B u_{L\to i} s_i \Big] = \sum_{s_j} \exp \bigg[ \B \Big(
(u_{\cP\to i} + u_{\cL\to i})\, s_i +
(U_{\cP\to L} + U_{\cL \to L} + J_{ij})\, s_i s_j + \nonumber\\
(u_{\cP\to j} + u_{\cL\to j} + u_{A\to j} + u_{B\to j} + u_{U\to j})\, s_j
\Big) \bigg]
\end{align}
where the indices refer to the notation used in
Fig. \ref{fig:message2D} and $J_{ij}$ is the interaction coupling
constant between spins $s_i$ and $s_j$.  This equation naturally
defines the updating rule for the message $u_{L\to i}$:
\begin{equation}
u_{L\to i} = \hat{u}(u_{\cP\to i} + u_{\cL \to i},\;
U_{\cP\to L} + U_{\cL\to L} + J_{ij},\;
u_{\cP\to j} + u_{\cL\to j} + u_{A\to j} + u_{B\to j} + u_{U\to j})\;,
\label{eq:message-u}
\end{equation}
where
\[
\hat{u}(u,U,h) \equiv u + \frac{1}{2\B} \log\frac{\cosh\B(U+h)}{\cosh\B(U-h)}
\]
Note that the usual cavity equation for fields in the Bethe
approximation \cite{basicBethe} is recovered if all contributions
from plaquettes $\cP$ and $\cL$ are set to zero.

Working in a similar way for the second equation in
(\ref{eq:message2D}) we end up with the updating rule for the message
$(U_{\cP\to L},u_{\cP\to i},u_{\cP\to j})$ sent from any given
plaquette region $\cP$ to one of its children links $L$ (see right
picture in Fig.~\ref{fig:message2D})
\begin{eqnarray}
U_{\cP\to L} &=& \frac{1}{4\B}\log\frac{K(1,1)K(-1,-1)}{K(1,-1)K(-1,1)} \nonumber \\
u_{\cP\to i} &=& u_{\cD\to i} - u_{D\to i} + \frac{1}{4\B}\log\frac{K(1,1)K(1,-1)}{K(-1,1)K(-1,-1)}
\label{eq:message-Uuu} \\
u_{\cP\to j} &=& u_{\cU\to j} - u_{U\to j} + \frac{1}{4\B}\log\frac{K(1,1)K(-1,1)}{K(1,-1)K(-1,-1)} \nonumber
\end{eqnarray}
where
\begin{eqnarray*}
K(s_i,s_j) &=& \sum_{s_k,s_l} \exp \bigg[
\B \Big( (U_{\cU\to U} + J_{jk}) s_j s_k +
(U_{\cR\to R} + J_{kl}) s_k s_l + (U_{\cD\to D} + J_{li}) s_l s_i + \\
&& (u_{U\to k} + u_{C\to k} + u_{E\to k} + u_{R\to k}) s_k +
(u_{R\to l} + u_{F\to l} + u_{G\to l} + u_{D\to l}) s_l \Big) \bigg]
\end{eqnarray*}

Equations (\ref{eq:message-u}) and (\ref{eq:message-Uuu}) are
equivalent to equations in (\ref{eq:message2D}), once multipliers
(messages) are parametrized in terms of fields. For instance, note
that the $\mu$ multipliers in the left hand side of second equation in
(\ref{eq:message2D}) appear now subtracted in the right hand side of
\rref{eq:message-Uuu}.

The field notation is more comprehensible and has a clear physical
meaning. Each plaquette $\cP$ is telling its children links $L$ that
they should add an effective interaction term $U_{\cP\to L}$ to the
real interaction $\Jij$, due to the fact that spins $s_i$ and $s_j$
are also interacting through the other three links in the plaquette
$\cP$. Fields $u$ act like magnetic fields upon spins, and the
complete $\nu_{\cP\to L}(s_i,s_j)-$message is characterized by the
triplet $(U_{\cP\to L},u_{\cP\to i},u_{\cP\to j})$, and will be
referred from now on as $Uuu-$message. Furthermore, it is clear that
some fields enter directly in the message-passing equations like
$u_{\cP\to i}$ and $u_{\cL\to i}$ in \rref{eq:message-u} and
$u_{\cD\to i}$ and $u_{\cU\to j}$ in \rref{eq:message-Uuu}. Also note
that since our model has no external field, the fields $u$ break the
symmetry of the original Hamiltonian whenever they are non zero. For
instance, in the ferromagnetic model, when all $\Jij=J$, these fields
are zero at high temperature and become non zero at Kikuchi's critical
temperature $T=1.42$ \cite{kikuchi}, implying a spontaneous
magnetization in the ferromagnet.

\section{The Dual approximation for the paramagnetic phase}
\label{Dual}

Unfortunately, the iterative message-passing algorithm for solving the
GBP equations (\ref{eq:message-u}) and (\ref{eq:message-Uuu}) often
does not converge on finite dimensional lattices.  While this is
expected if long range correlations are present, it is rather
disappointing that it happens also in the paramagnetic phase, where
one would like to find easily the solution to the model.  Here we are
going to focus only on the paramagnetic phase, and propose an improved
solving algorithm based on physical assumptions.

In the paramagnetic phase of any spin model defined by the Hamiltonian
in \rref{eq:Ham}, that is with no external field, variables have no
polarization or magnetization: this in turn implies that in the
solution all $u-$fields must be zero, and only $U-$fields should be
fixed self-consistently to non zero values.

This paramagnetic solution has some interesting properties. First, it
is always a solution of the GBP equations, since \rref{eq:message-u}
and \rref{eq:message-Uuu} are self-consistent with all $u=0$. This
means that starting from unbiased messages (all $u=0$) the iterative
GBP algorithm keeps this property. Second, the paramagnetic Ansatz is
correct, from the GBP perspective, at least at high enough
temperatures, meaning that even if we start with biased messages ($u
\neq 0$), the iterative algorithm converges to all $u=0$ at high
temperatures. And last, but not least, the well studied physical
behavior for the 2D EA model with zero-mean random interactions
$\Jij$, is expected to remain always paramagnetic, \ie to have no
transition to a spin glass phase at any finite temperature
\cite{JLMM}.  Therefore, the Ansatz $u=0$ is both physically plausible
and algorithmically desirable.

\begin{figure}[htb]
\begin{center}              
\includegraphics[width=0.5\textwidth]{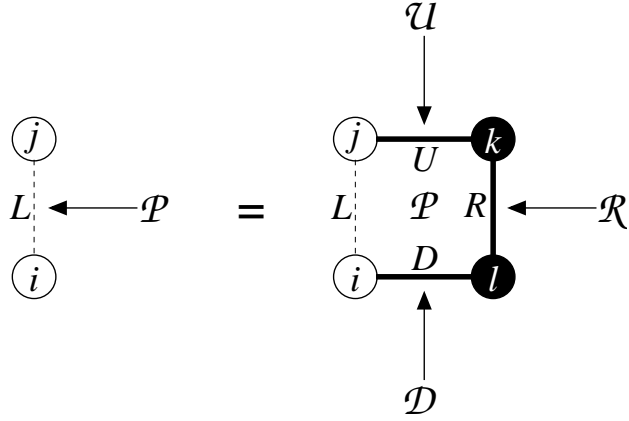} 
\end{center} 
\caption{Message passing of correlation messages in the dual
  approximation. In the right hand side the trace is taken over the
  black spins.}
\label{fig:messageDual}
\end{figure}

Under the paramagnetic Ansatz, which we shall also call Dual
approximation for a reason to be explained soon, the message-passing
equation (\ref{eq:message-u}) is irrelevant, as it is always satisfied
given that $\hat{u}(0,U,0)=0$, while \rref{eq:message-Uuu} now turns
into (see Fig.~\ref{fig:messageDual})
\begin{multline}
U_{\cP\to L} = \widehat{U}(U_{\cU\to U},U_{\cR\to R},U_{\cD\to D}) =\\
\frac{1}{\B}\:\arctanh\Big[\tanh \B (U_{\cU\to U} + J_{jk})
\tanh \B (U_{\cR\to R} + J_{kl}) \tanh \B (U_{\cD\to D} + J_{li}) \Big]\;.
\label{eq:message-U}
\end{multline}
The only relevant messages now are those associated to the multipliers
$\nu_{\cP\to L}(s_i,s_j) = \B U_{\cP\to L}\, s_i\, s_j$, and they will
be refereed to as $U-$messages. \rref{eq:message-U} can be interpreted
as a correlation message-passing equation, giving the new interaction
field $\widehat{U}$ that a certain link shall experience as a
consequence of the correlations transmitted around the plaquette. The
belief \rref{eq:beliefs} also simplify. Obviously $b(s_i)=0.5$ for
every spin in the graph, and the link and plaquette beliefs are
\begin{eqnarray}
b_L(s_i,s_j) &=& \frac{1}{Z_L} e^{\B (U_{\cL\to L} + U_{\cP\to L} +
  J_{ij}) s_i s_j}\;, \label{eq:belief_U}\\
b_\cP(s_i,s_j,s_k,s_l) &=& \frac{1}{Z_\cP} e^{\B (U_{\cL\to L} + J_{ij})
  s_i s_j + \B (U_{\cU\to U} + J_{jk}) s_j s_k + \B (U_{\cR\to R} + J_{kl})
  s_k s_l + \B (U_{\cD\to D} + J_{li}) s_l s_i}\;. \nonumber
\end{eqnarray}

The Dual algorithm we are proposing to study the paramagnetic phase of
the EA model, is a standard message passing algorithm for the
$U$-messages, which works as follows.

\algsetup{indent=4em}
\begin{algorithmic}[1]
\STATE Start with all $U$-messages null
\REPEAT
\STATE Choose randomly one plaquette $\cP$ and one of its children
links $L$
\STATE Update the field $U_{\cP\to L}$ according to \rref{eq:message-U}
as in Fig.~\ref{fig:messageDual}
\UNTIL{The last change for any $U$-message is less than $\epsilon$ (we
  use typically $\epsilon = 10^{-10}$)}
\RETURN The beliefs $b_L(s_i,s_j)$ defined in \rref{eq:belief_U} for
every pair of neighboring spins
\end{algorithmic}
Some damping factor $\gamma \in [0,1)$ can be added in the update step
$U_{\cP\to L} = \gamma U_{\cP\to L} + (1-\gamma) \widehat{U}$ in order
to help convergence.

\subsection{Mapping to the dual model}

It is worth noticing that \rref{eq:message-U} is nothing but the
BP equation for the corresponding dual model
(hence the name of the algorithm).

The dual model has a binary variable $x_{ij} \equiv s_i s_j$ on every
link of the original model, and the original coupling constants play
now the role of an external polarizing (eventually random) field
\[
\Ham_\text{dual}(\vec x) = - \sum_{\< i,j \>} \Jij x_{ij}\;.
\]
This Hamiltonian looks like the sum of independent variables, but this
is not the case. The dual variables $x_{ij} = \pm 1$ must satisfy a
constraint for each cycle (or closed path) in the original graph,
enforcing that their product along the cycle must be equal to 1. On a
regular lattice any closed path can be expressed in terms of
elementary cycles of 4 links (the plaquettes) and so it is enough to
enforce the constraint on every plaquette: $x_{ij} x_{jk} x_{kl}
x_{li} = 1$. The Gibbs-Boltzmann probability distribution for the dual
model is then given by
\begin{equation}
P(\vec{x}) = \frac{1}{Z} e^{-\B \Ham_\text{dual}(\vec x)}
\prod_{\langle i,j,k,l \rangle} \delta_{x_{ij} x_{jk} x_{kl}
  x_{li},\, 1}\;,
\label{eq:measureDual}
\end{equation}
where the product runs over all elementary plaquette.

The model described by the probability measure in
\rref{eq:measureDual} can be viewed as a constraint satisfaction
problem with a non uniform prior (given by $e^{-\B
  \Ham_\text{dual}(\vec x)}$). It is straightforward to derive the BP
equations for such a problem. Indeed by defining the marginal for the
variable $x_{ij}$ on link $L$ in the presence of the solely
neighboring plaquette $\cP$ as $\big(1+x_{ij}\tanh \B U_{\cP\to
  L}\big)/2 \propto \exp(\B U_{\cP\to L}x_{ij}\big)$, the BP equations
read
\begin{multline}
\frac12 \big(1+x_{ij}\tanh \B U_{\cP\to L}\big) \propto\\
\sum_{x_{jk},x_{kl},x_{li}}
e^{\B U_{\cU\to U}x_{jk}} e^{\B J_{jk} x_{jk}}
e^{\B U_{\cR\to R}x_{kl}} e^{\B J_{kl} x_{kl}}
e^{\B U_{\cD\to D}x_{li}} e^{\B J_{li} x_{li}}
\delta_{x_{jk}x_{kl}x_{li},\,x_{ij}} \propto\\
\mathop{\sum_{x_{jk},x_{kl},x_{li}:}}_{x_{jk}x_{kl}x_{li}=x_{ij}}
\!\!\!\!\!\!\!\!\!
\big(1 + x_{jk} \tanh \B (U_{\cU\to U} + J_{jk})\big)
\big(1 + x_{kl} \tanh \B (U_{\cR\to R} + J_{kl})\big)
\big(1 + x_{li} \tanh \B (U_{\cD\to D} + J_{li})\big)=\\
1+x_{ij}\tanh \B (U_{\cU\to U} + J_{jk})
\tanh \B (U_{\cR\to R} + J_{kl}) \tanh \B (U_{\cD\to D} + J_{li})\;.
\end{multline}
In the second summation the terms containing one or two $x$ variables
sums to zero, while the other two terms are those written in the last
expression. Equating the first and the last expressions, this equation
is manifestly equal to \rref{eq:message-U}.

\subsection{Average Case Solution}

GBP in general, and the Dual approximation in particular, are methods
for the study of the thermodynamical properties of a given problem.
However, in the limit of large systems ($N\to \infty$, thermodynamical
limit), we expect a typical behavior to arise. This is the so called
\textit{self-averaging} property of disordered systems. By typical we
mean that almost every realization of the interactions $\Jij$ will
result in a system whose thermodynamical properties (free energy,
energy, entropy) are very close to the average value.

Normally, in disordered systems, we cope with the $N\to \infty$ limit
and with the average over the random $\Jij$ by the replica method.
The application of the replica trick to regions graph approximations
is a challenging task \cite{TLMF10}. However, we can still grasp the
average case behavior with a cavity average case solution of the dual
message-passing equations, at the price of neglecting the local
structure of the graph (beyond plaquettes).

The idea is to represent the set of $U$-messages flowing in any given
graph, by a population of messages $Q(U)$. Then the message-passing
\rref{eq:message-U} is used to obtain such population in a
self-consistent way. More precisely, in every iteration three messages
$U_1,U_2,U_3$ are randomly drawn from the population $Q(U)$ and a new
message $U_0=\widehat{U}(U_1,U_2,U_3)$ is computed by
\rref{eq:message-U} using three couplings randomly selected from
$P(J)$.  The obtained message $U_0$ is put back into the population,
and the iteration is repeated many times, until the population
stabilizes.

Once we have the self consistent population of messages, we can
compute the average energy
\begin{equation}
E_{\mbox{Ave}} = \<-\Jij \tanh\B\left( \Jij +U_1 +U_2 \right)
\>_{Q(U_1),Q(U_2),P(\Jij)}
\label{eq:aveEner}
\end{equation}
by a random sampling of the population and of the interactions. The
average case solution is supposed to be very good whenever the network
of interactions has no or few short loops. This is not the case in any
finite dimensional lattice, since there the short loops (plaquettes)
are abundant.  Nonetheless, the average case solution gives a
reasonably good approximation to the single instance results in 2D and
3D, as shown in the next Section.

\section{Results on 2D EA model}
\label{Results2D}

Message-passing algorithms work fine in the high temperature regime
($T > T_c$) of models defined on random topologies: this is the reason
why these methods have been successfully applied in random constraint
satisfaction problems, like random-SAT or random-Coloring
\cite{KSAT,Col}. However, when used on regular finite-dimensional
lattices, they can experience difficulties even in the paramagnetic
phase, because the presence of short loops spoils message-passing
convergence.

\begin{figure}[!htb]
\includegraphics[angle=270,width=0.8\textwidth]{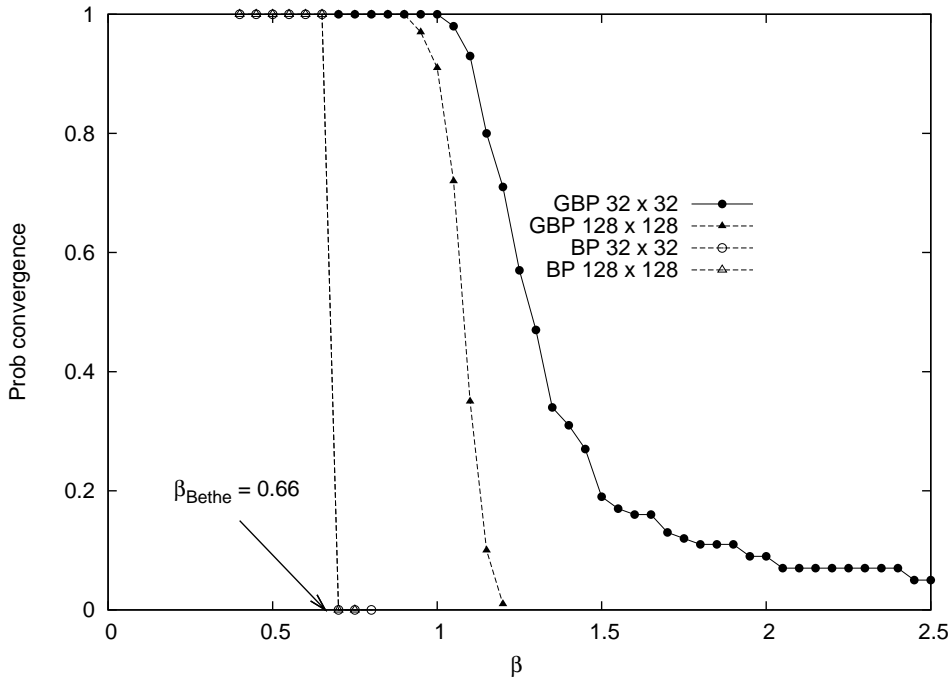}
\caption[0]{Convergence probability of BP (Bethe approximation) and
  GBP on a 2D square lattice, as a function of inverse
  temperature. Data points are averages over 100 systems with random
  bimodal interactions. System sizes are $N=L^2$ with $L=32, 128$ and
  a damping factor $\gamma=0.5$ has been used in the iteration of the
  message-passing equations. The Bethe spin glass transition is
  expected to occur at $\beta_\text{Bethe} \simeq 0.66$
  ($T_\text{Bethe}\simeq 1.52$) for a random graph with the same
  connectivity as the 2D square lattice. Notably, that temperature
  also marks the convergence threshold for BP equations in the 2D
  square lattice. GBP, on the contrary, reaches lower temperatures,
  but eventually stop converging.}
\label{fig:betheEA2D}
\end{figure}

It is well known that on a random graph of fixed degree (connectivity)
$c=4$ the cavity approximation gives a paramagnetic result above
$T_{\text{Bethe}} \simeq 1.52$ (i.e.\ $\beta_\text{Bethe} \simeq
0.66$) with all cavity fields $u_i=0$.  Below the Bethe critical
temperature, this solution becomes unstable to perturbations, and we
expect many solutions to appear with non trivial messages $u_i\neq
0$. The presence of many solutions in the messages passing equations
is connected to the existence of many thermodynamical states in the
Gibbs-Boltzmann measure, or, equivalently, to the presence of replica
symmetry breaking. The appearance of such a spin glass phase is also
responsible for the lack of convergence of message-passing equations,
since the intrinsic locality of the message-passing equations fails to
coordinate distant regions of the graph (which are now long-range
correlated). As a consequence, the application of BP to the 2D EA
model (that also has fixed degree $c=4$) still finds the paramagnetic
phase at high temperatures, but below $T_{\text{Bethe}}$, the Bethe
instability takes the message-passing iteration away from the $u=0$
solution and does not allow the messages to convergence to a fixed
point (\ie the algorithm wanders forever).  In Figure
\ref{fig:betheEA2D} we show the convergence probability for the BP
message-passing equations in the 2D EA model.

On the other hand, a straightforward GBP Parent-to-Child
implementation does not fully overcome this problem. At high
temperatures, the Parent-to-Child equations converge to a paramagnetic
solution with all $u=0$ and non trivial $U\neq0$, which turns out to
be the same solution found by our Dual algorithm. When going down in
temperature, the convergence properties of the algorithm worsen, and
are sensitive to tricks like damping and bounding in the fields. A
thorough discussion of these properties is left for a future work, but
let us summarize that typically the algorithm stop converging at low
temperatures, somewhere below $T_\text{Bethe}$, as shown in Figure
\ref{fig:betheEA2D}.

So, in general, BP and GBP equations are not simple to use in finite
dimensional systems at low enough temperatures: this warning was
already reported in Refs.~\cite{pelizzola,HAK03,MooKap07}. Indeed a
different method for extremizing the constrained free energy named
Double Loop algorithm \cite{HAK03,Yuille} was developed to overcome
such difficulties. As mentioned earlier, Double Loop guarantees
convergence of the beliefs, on any topology, with or without short
loops.  Given the convergence problems in GBP, researchers typically
resort to Double Loop algorithms to extremize region graph
approximations to the free energy, below the Bethe critical
temperature.

In order to make a fair comparison with our Dual algorithm, we have
used an optimized code for GBP and Double Loop algorithms: the open
source LibDai library written in C++ \cite{libdai0.2.3}.

The first interesting result of our work is that our Dual algorithm
converges at all temperatures, just as
Double Loop does. The reason why it converges is that there are no
$u$-messages, so the Bethe instability will not affect our
message-passing iteration.

The second relevant result of our Dual algorithm, is the fact that it
finds the same solution found by the Double Loop algorithm at all
temperatures. In other words, the direct extremization of the region
graph approximation to free energy \rref{eq:freeen} via a Double Loop
algorithm finds a paramagnetic solution characterized by the beliefs
$b_i(s_i)=0.5$ and $b_L(s_i,s_j) = \frac{1}{z} e^{-\B \tilde{J}_{ij}
  s_i s_j}$; and the effective interactions $\tilde{J}_{ij}$ found by
the Double Loop algorithm are exactly equal to those found with our
Dual algorithm, $\tilde{J}_{ij} = \Jij + U_{\cP \to L} + U_{\cL \to
  L}$. This means that beliefs and correlations found by the two
algorithms are identical: $\< s_i s_j \>_\text{Double Loop} = \< s_i
s_j \>_\text{Dual}$.

\begin{figure}[!htb]
\includegraphics[angle=270,width=0.8\textwidth]{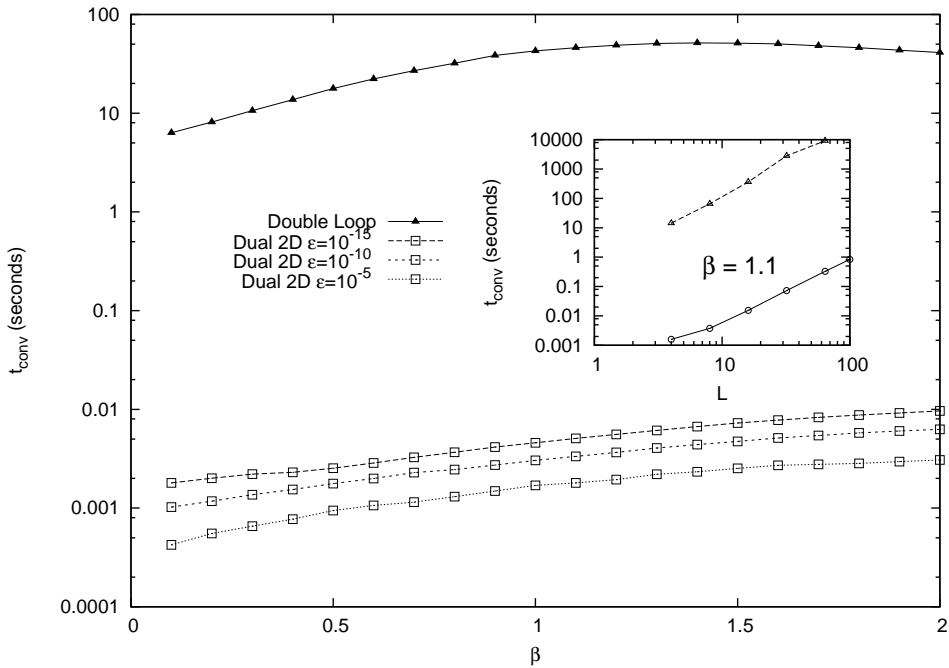}
\caption[0]{Running times of the Double Loop algorithm
  \cite{HAK03,libdai0.2.3} (libdai) and the Dual algorithm averaged
  over 10 realizations of a 2D $8\times8$ EA model with Gaussian
  interactions. Generally the Double Loop algorithm requires a time 4
  orders of magnitude larger than that used by the Dual algorithm.
  Three different precision goals where used for the Dual algorithm
  $10^{-5},10^{-10},10^{-15}$, while the precision of the Double Loop
  algorithm is $10^{-9}$. The inset shows the behavior of the running
  times for both algorithm versus the system size $L=\sqrt{N}$.}
\label{fig:timeDAIDual}
\end{figure}

The third result is that the running times of our Dual algorithm are
nearly four orders of magnitude smaller than those required by the
Double Loop implementation in libdai, at least in a wide range of
temperatures (see figure \ref{fig:timeDAIDual}). More precisely, the
convergence time of the Dual algorithm growth exponentially with
$\B=1/T$, but still, in the relevant range of temperatures where the
region graph approximation is a good approximation (not too low
temperatures), the running time is always roughly a factor $10^4$
smaller than Double Loop.

\subsection{Dual approximation \textit{vs} Monte Carlo simulations}

\begin{figure}[!htb]
\includegraphics[angle=270,width=0.8\textwidth]{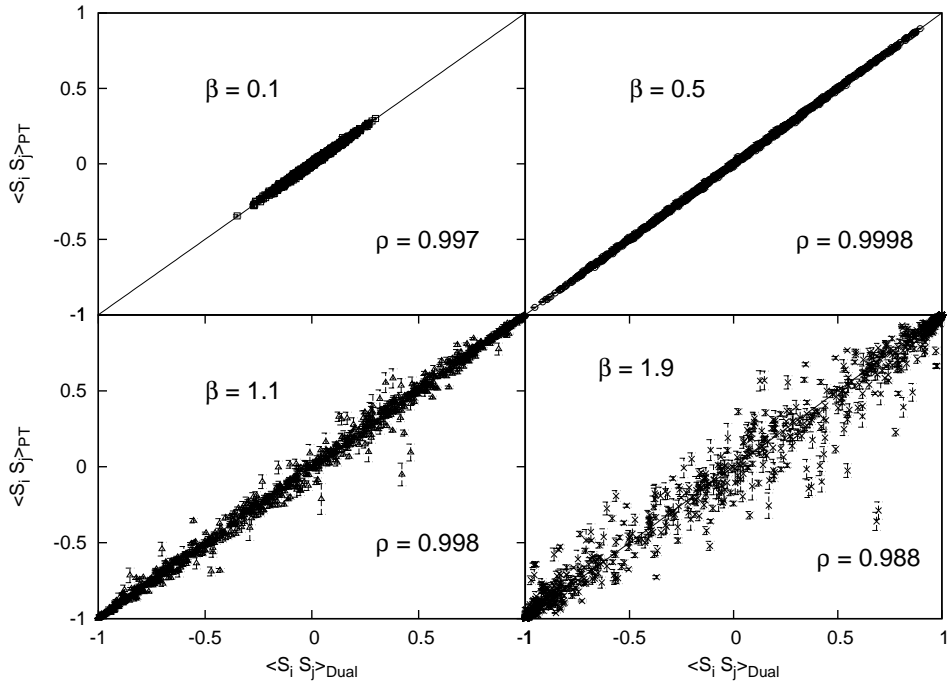}
\caption[0]{Comparison between the correlations $\< s_i s_j
  \>_\text{Dual}$ obtained by the Dual algorithm and the nearly exact
  correlations obtained by a Parallel Tempering simulation. We used a
  $64\times 64$ EA model with Gaussian interactions. At each
  temperature the data correlation coefficient $\rho$ is reported.}
\label{fig:PT-Dual16}
\end{figure}

The fact that our Dual algorithm provides the same results (and much
faster) than the Double Loop algorithm is a very good
news. Essentially is telling us that we are not loosing anything by
restricting the space of possible messages, as far as the region graph
approximation is concerned. However, the ultimate comparison for the
approximation has to be done with the exact marginals and
correlations. In figure \ref{fig:PT-Dual16} we show a comparison
between the exact correlations $\< s_i s_j \>_\text{PT}$ of
neighboring spins obtained with a Parallel Tempering (PT) Monte Carlo
simulation, and the Dual approximation estimate for the same two-spins
correlations.  The coincidence between $\< s_i s_j \>_\text{PT}$ and
$\< s_i s_j \>_\text{Dual}$ is essentially perfect at high
temperatures, and it becomes weaker as the temperature is decreased.
The reason for the discrepancies is obviously the fact that we are
using an approximation in which collective behaviors of spins is
accounted exactly only until the plaquette level; more distant
correlations are approximated and these correlations become more
important at low temperatures.

\begin{figure}[!htb]
\includegraphics[angle=270,width=0.8\textwidth]{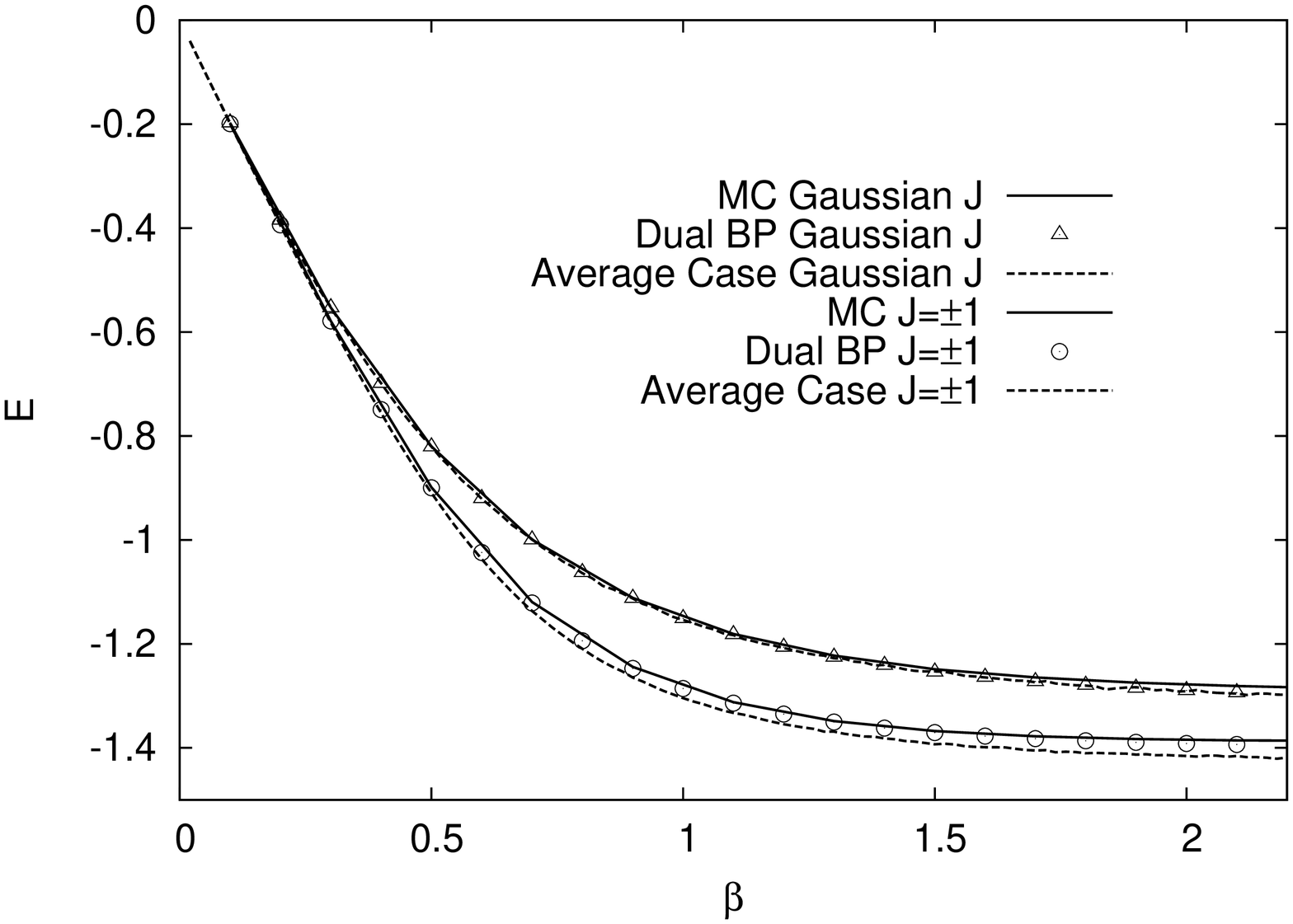}
\caption[0]{Energy as a function of the inverse temperature $\B$ for a
  $64\times64$ 2D EA model, with both types of interactions, Gaussian
  and bimodal. Full lines represent the exact thermodynamical energy
  as obtained by a Monte Carlo simulation, points are the energies
  obtained under the Dual approximation, and dashed lines are the
  average case energies.}
\label{fig:ene2D}
\end{figure}

Given such a good correspondence between the correlations under the
Dual approximation and the true correlations, we expect a very good
estimate for the energy too. In Figure~\ref{fig:ene2D} we show with
points the energy under the Dual approximation and with full lines the
Monte Carlo exact energy: the data are indeed very close.  The dashed
lines show the average case energy for the Dual approximation,
\rref{eq:aveEner}. In spite of the fact that the average case does not
take into account the local structure of the lattice, the average case
energy is quite close to the single instance one.

\subsection{Ground State Configuration in 2D}
\label{GroundState}

\begin{figure}[!htb]
\includegraphics[angle=0,width=0.4\textwidth]{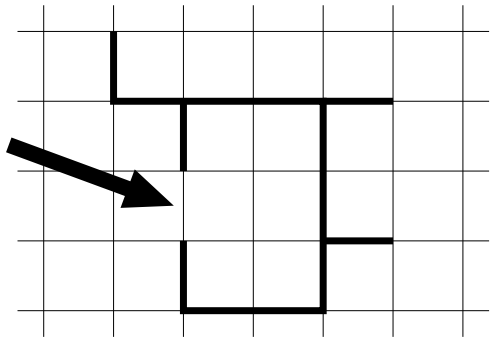}
\caption[0]{Even if the link marked by the arrow is not the most
  polarized link according to the marginals provided by the Dual
  algorithm, the spins it connects are fully correlated by the fact
  that they belong to a cluster of frozen links (bold lines).
  Therefore, the marked link must be immediately fixed accordingly.}
\label{fig:cluster}
\end{figure}

The good agreement between the correlations found by the Dual
algorithm, and those found in a Monte Carlo simulation, for the 2D EA
model, compels us to push this correspondence down to $T=0$. More
precisely, using the correlations obtained by our Dual algorithm at
low temperatures, we try to compute a ground state configuration by
the following procedure. The idea is to freeze iteratively the
relative position $s_i s_j$ of those interacting spins that are more
strongly correlated, which is done by setting $J_{ij}\to \pm\infty$,
and re-running the Dual algorithm until convergence every time one
pair of spins is frozen. Note that freezing the relative position of
spins is equivalent to freeze the dual variable $x_{ij} = s_i s_j$.
The freezing procedure is very simple, but for the fact one has to
check that frozen links must be consistent with a spin
configuration. More precisely, frozen $x_{ij}$ variables must satisfy
the requirement that on any closed loops the product is one,
\begin{equation}
\prod_{ij}^{\text{closed loop}} x_{ij} = 1\;.
\label{eq:condition}
\end{equation}
For very short loops the satisfaction of this condition is
automatically induced by the Dual algorithm: for example if three
links on a plaquette freeze, the fourth link is immediately frozen to
a value satisfy condition in \rref{eq:condition}.  However, for longer
loops (as the one shown in Fig.~\ref{fig:cluster}), the propagation of
these constraints by the Dual algorithm is not perfect, since the
information degrades with distance beyond the plaquette level.  Then
we need to enforce the constraints of \rref{eq:condition} by a proper
algorithm.  At each stage of the freezing process, we define the
clusters of frozen links as follows: if two frozen links share a spin,
then they belong to the same cluster. In Figure~\ref{fig:cluster} a
cluster of frozen links is represented by bold lines. Notice that,
once a spin is fixed in a cluster, all other spins are fixed as well
by the frozen correlations. On the other hand, different clusters of
spins can have arbitrary relative orientations.

Consider now the situation depicted in Figure~\ref{fig:cluster} and
focus on the value of the correlation between the two spins connected
by the link marked by an arrow. From the fact these two spins belong
to the same cluster of frozen links (shown as bold link in
Figure~\ref{fig:cluster}) we know they are perfectly correlated,
however by running the Dual algorithm we could get a weak value for
this correlation and then proceed by freezing a different link. A set
of sub-optimal choices of this kind may finally produce a
configuration of frozen links where the constraints in
\rref{eq:condition} are not all satisfied.  In order to avoid these
constraint violations we force any link whose spins are already part
of the same cluster to be polarized accordingly. The freezing
algorithm, therefore, works as follows.

\label{alg:reinforcement}
\algsetup{indent=4em}
\begin{algorithmic}[1]
\REPEAT
\STATE run the Dual algorithm until convergence (at a low enough
temperature) 
\STATE find the link $L$ with largest finite
$\tilde{J}_L=J_L+U_{\cP\to L}+U_{\cL\to L}$ 
\STATE freeze that link by setting $J_L \leftarrow
\Sign(\tilde{J}_L)\, \infty$ 
\IF {link $L$ is connected to clusters $\mathbf{C}$ and $\mathbf{C'}$
  of frozen links} 
\STATE merge clusters $\mathbf{C}$, $\mathbf{C'}$ and link $L$ in a
unique cluster 
\ELSE
\IF {link $L$ is connected to a single cluster $\mathbf{C}$ of frozen
  links} 
\STATE add link $L$ to cluster $\mathbf{C}$
\ELSE  
\STATE create a new cluster with link $L$
\ENDIF
\ENDIF
\FORALL {non-frozen links $L'$ at the boundaries of a cluster of
  frozen links} 
\IF {link $L'$ shares both spins with the same cluster} 
\STATE freeze link $L'$ accordingly \COMMENT{To avoid violations of
  constraint} 
\ENDIF
\ENDFOR 
\UNTIL{all links are frozen}
\RETURN the spins configuration obtained by setting one spin and
fixing the rest according to frozen links
\end{algorithmic}

\begin{figure}[!htb]
\includegraphics[angle=270,width=0.8\textwidth]{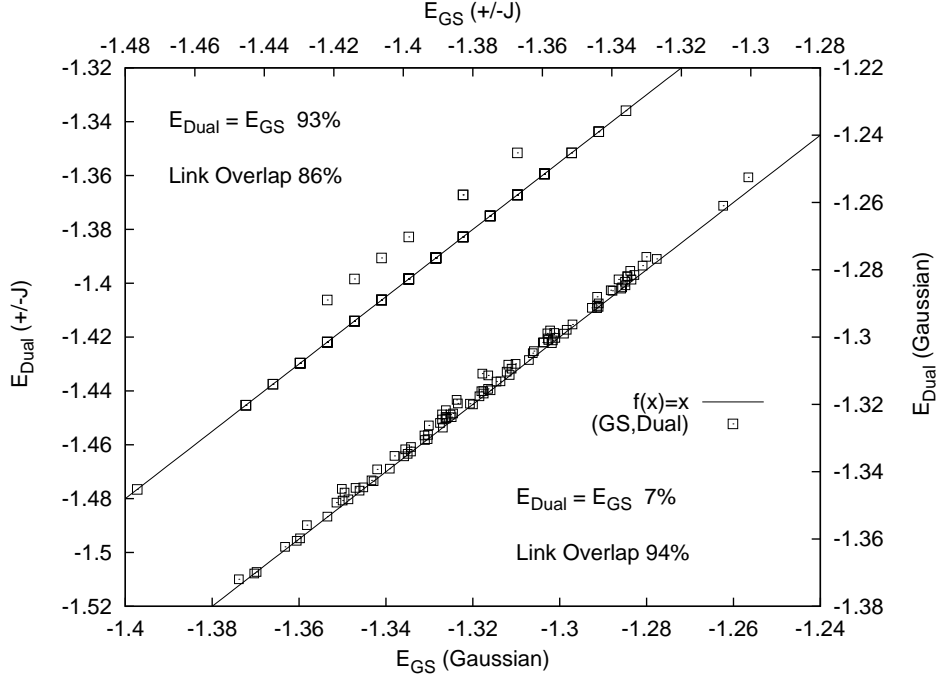}
\caption[0]{Correlation of the Dual+Freezing ground state energy with
  the exact ground state energy in $N=16\times 16$ systems. The top
  left points correspond to 100 bimodal systems $\Jij = \pm1$, while
  the right bottom points correspond to 100 systems with Gaussian
  interactions. For bimodal interactions, the degeneracy of the ground
  state improves the probability of actually finding the correct
  ground state energy ($93\%$), and conversely reduces the expected
  link correlation overlap with the exact ground state solution
  ($86\%$). For Gaussian interactions, the ground state is not
  degenerated, and only in $\sim 7\%$ of the samples the Dual+Freezing
  method finds the actual ground state; however the average link
  overlap is very high ($94\%$). The line $f(x)=x$ is shown to guide
  the eye. Kindly note that two set of axes are being used.}
\label{fig:EGSvsEBP}
\end{figure}

The results obtained with this freezing procedure are quite good. In
figure \ref{fig:EGSvsEBP} we compare the resulting ground state
energies with the exact solutions obtained using a web service running
an exact solving algorithm~\cite{juergengroundwww}. We used an
ensemble of 100 EA models on the 2D square lattice with Gaussian
interactions (so the ground state is not degenerate) and with bimodal
interactions.  Most points are along the bisecting line, meaning that
the ground state found by either methods are the same. The relative
error for the ground state energy is $0.0013$ for Gaussian systems,
and $0.00078$ for bimodal systems. Looking at how many links are
frustrated in one of the solutions and unfrustrated in the other, we
found that the Dual+Freezing algorithm returns $94.3\%$ of the correct
link correlation signs with respect to the true ground state solution
for the Gaussian models. For the bimodal system, given the degeneracy
of the ground state configuration, it is more probable to find the
actual ground state energy but for the same reason the link overlap
between the exact Ground State and the one found with Dual+Freezing is
significantly lower ($86.0\%$).
 
In general we believe these results on the ground states to be very
encouraging, considering that the Dual algorithm is very fast and not
restricted to the 2D case (at variance to fast exact algorithms for
computing ground states). They provide evidence that the marginals
obtained by this Dual algorithm are reliable even at very low
temperatures.

\section{Generalization to other dimensions}
\label{Results3D}

Let us now consider the region graph based approximation to the free
energy for a generic $D$-dimensional (hyper-)cubic lattice, using the
same hierarchy of regions: square plaquettes, links and spins. After
computing the counting numbers for a general $D$-dimensional lattice,
see \rref{eq:countingnumbers}, the free energy approximation becomes
\begin{eqnarray}
\B F &=& \sum_\cP \sum_{\sigma_\cP} b_\cP(\sigma_\cP) \log
\frac{b_\cP(\sigma_\cP)}{\exp(-\B E_\cP(\sigma_\cP))}
\qquad\text{Plaquettes} \nonumber\\ 
& & -(2D-3) \sum_{L} \sum_{\sigma_L} b_L(\sigma_L) \log
\frac{b_L(\sigma_L)}{\exp(-\B E_L(\sigma_L))}
\qquad\text{Links}  \label{eq:freeen_D_dim} \\ 
& & +(2 D^2-4 D+ 1)\sum_{i} \sum_{s_i} b_i(s_i) \log
\frac{b_i(s_i)}{\exp(-\B E_i(s_i))}  \qquad\mbox{Spins} \nonumber 
\end{eqnarray}
Plaquettes are still the biggest regions considered at so have
counting number 1, but now each link is contained in $2(D-1)$
plaquettes, and each spin is in $2 D$ links and $2 D (D-1)$
plaquettes.  The message passing equations for the Dual algorithm in
$D$ dimensions are then
\begin{multline}
U_{\cP\to L} = \frac{1}{\B}\:\arctanh\Bigg[\tanh \B
\left(\sum_i^{2(D-1)-1} U_{\cU_i \to U} + J_U\right)\\ 
\tanh \B \left(\sum_i^{2(D-1)-1} U_{\cR_i \to R} + J_R\right) \tanh \B
\left(\sum_i^{2(D-1)-1} U_{\cD_i \to D} + J_D\right) \Bigg]\;, 
\end{multline}
where $\cU_i$ (resp.\ $\cR_i$ and $\cD_i$) are the $2(D-1)-1$
plaquettes containing the link $U$ (resp.\ $R$ and $D$) excluding
plaquette $\cP$.

In the high temperature phase, this Dual approximation with all $u=0$
should be still a valid approach for any dimensionality $D$.  At low
temperatures, however, the EA model in more than two dimensions have a
spin glass phase transition and, therefore, we expect the Dual
approximation to become poorer, as it can not account for a non
trivial order parameter.

By running the Dual algorithm for the 3D EA model we have found a
divergence of $U$-fields around $\B\simeq 0.39$ for bimodal couplings
and around $\B\simeq 0.41$ for Gaussian couplings. This divergence is
due to the fact the $U$-fields get too much self-reinforced under
iteration. This divergence does not come as a surprise given that it
happens also when studying the simpler pure ferromagnetic Ising
model. However in the ferromagnetic model the temperature at which
$U$-fields diverge is always below the critical temperature and so the
Dual algorithm still provides a very good description of the entire
paramagnetic phase.

Unfortunately in the 3D EA the divergence of $U$-fields takes place
well above the critical temperature (which is $T_c \simeq 1.12$ for
bimodal coupling and at $T_c \simeq 0.95$ for Gaussian couplings, see
Ref.~\onlinecite{KKY} for a summary of critical temperatures in 3D
spin glasses) and this would make the Dual algorithm of very little
use.  We have studied the origin of this divergence and we have found
a general principle for reducing the divergence of $U$-fields due to
self-reinforcement, thus improving the convergence properties of the
Dual algorithm.  The idea is the following.  When writing the Dual
approximation as a constraint satisfaction problem with a non uniform
prior, see \rref{eq:measureDual}, the constraints may be redundant.
This is the case for the 3D cubic lattice: indeed both the number of
links (i.e.\ variables in the dual problem) and the number of
plaquettes (i.e.\ constraints in the dual problem) are $3N$. So, if
constraints were independent, the entropy would be null at $\B=0$ and
negative for $\B>0$ (and this is clearly absurd).  The solution to the
apparent paradox is that constraints are not independent: actually
only $2/3$ of these are independent, and the remaining third is
uniquely fixed by the value of the former. In this way the correct
entropy is recovered at $\B=0$, given that a problem with $3N$
unbiased binary variables subject to $2N$ independent parity-check
constraints has entropy $N \log(2)$.  The dependence among constraints
can be easily appreciated by looking at the 6 plaquette around a cube:
if 5 of the 6 constraints are satisfied, then the sixth one is
automatically satisfied and redundant.

The general rule for improving the convergence of the Dual algorithm
is to remove redundant constraints (this principle is similar to the
maxent-normal property of region based free energy approximations
\cite{yedidia}).  Redundant constraints have no role in determining
the fixed point values for the beliefs (since they are redundant), but
during the iterations they provide larger fluctuations to messages and
may be responsible for the lack of convergence.  In practice, on a 3D
cubic lattice, we may remove redundant constraints in many different
ways: the basic rule states that one constraint (i.e.\ a plaquette)
should be removed for each elementary cube, otherwise if a cube
remains with its 6 plaquettes at least one redundant constraint will
exist. We have used two different ways of removing one constraint per
cube and we have found the same results in the entire paramagnetic
phase. So, for simplicity, we are going to present data obtained by
removing all constraints corresponding to plaquettes in the $xy$
plane.

\begin{figure}[!htb]
\includegraphics[angle=270,width=0.8\textwidth]{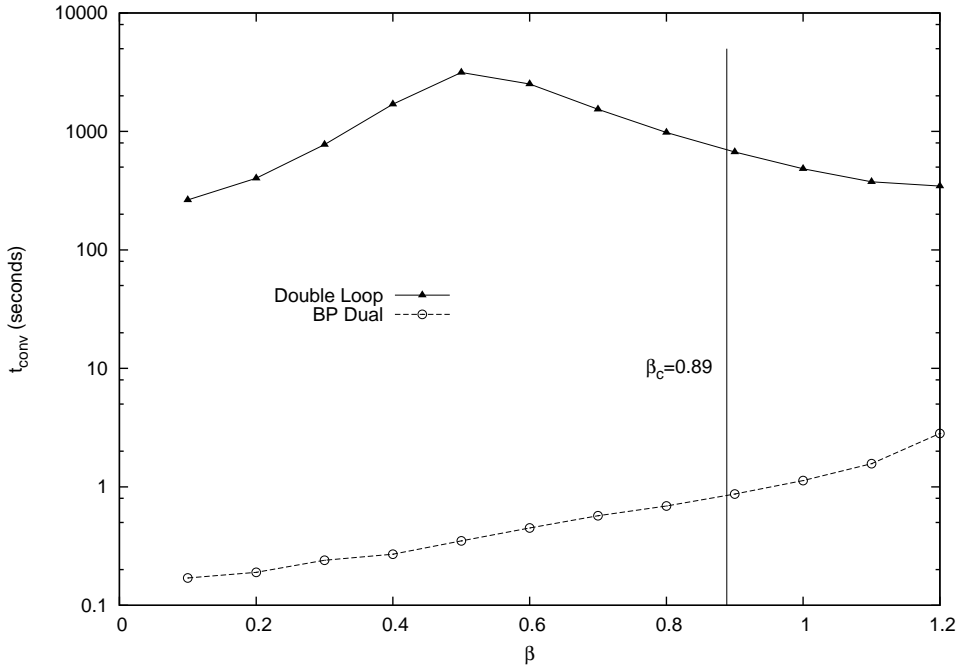}
\caption[0]{Running times of the Double Loop algorithm
  \cite{HAK03,libdai0.2.3} (libdai) and of the Dual algorithm on a
  $8\times8\times8$ EA model with bimodal interactions ($J_{ij}=\pm
  1$). The Dual algorithm is generally several orders of magnitude
  faster and returns the same solution as the Double Loop algorithm.}
\label{fig:timeDAIDual3D}
\end{figure}

\begin{figure}[!htb]
\includegraphics[angle=270,width=0.8\textwidth]{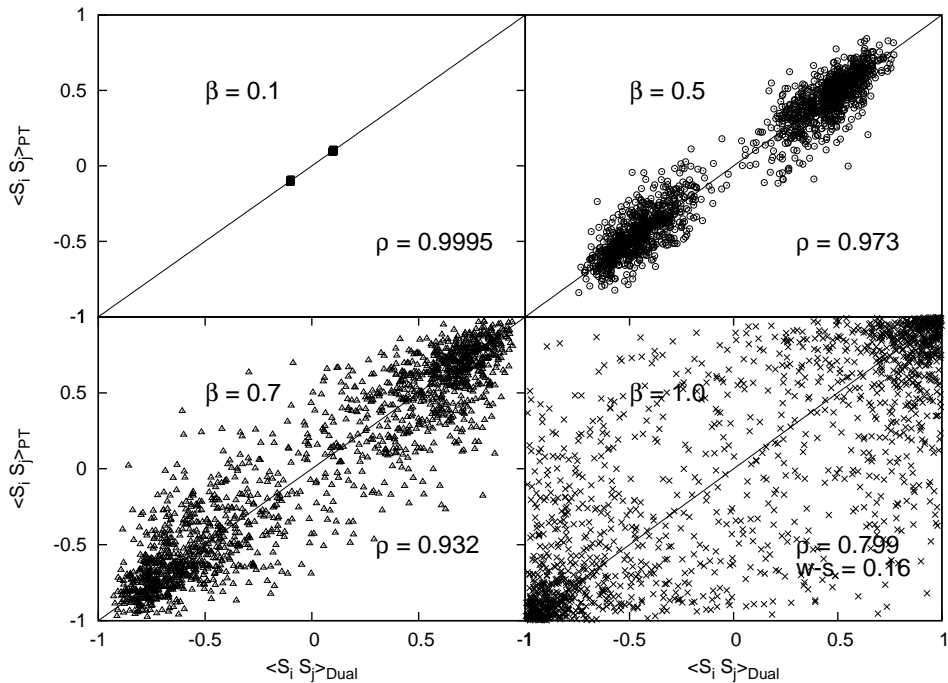}
\caption[0]{Comparison between the correlations $\< s_i s_j
  \>_\text{Dual}$ obtained with the Dual algorithm and the (nearly)
  exact correlations $\< s_i s_j \>_\text{PT}$ obtained with a
  Parallel Tempering simulation in a 3D EA model of size $8\times
  8\times 8$ with random bimodal interactions $\Jij=\pm1$. At each
  temperature the correlation coefficient $\rho$ is reported. For the
  lowest temperature shown, $\B =1.0$, we also report the fraction w-s
  of pairs of spins such that $\< s_i s_j \>_\text{Dual}\< s_i s_j
  \>_\text{PT}<0$.}
\label{fig:PT-Dual3D}
\end{figure}

\begin{figure}[!htb]
\includegraphics[angle=270,width=0.8\textwidth]{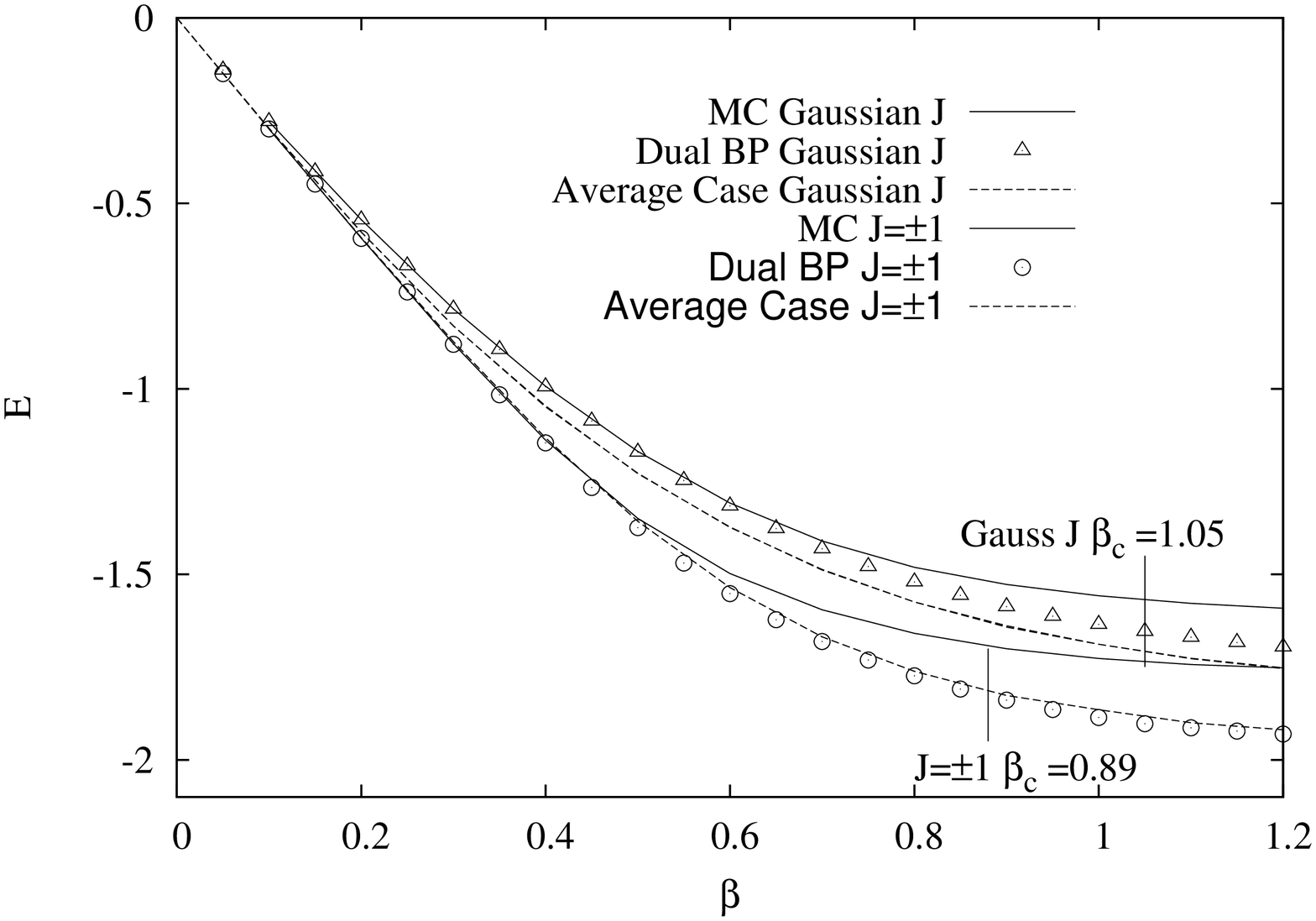}
\caption[0]{The energy predicted by the Dual approximation in 3D EA
  model, compared to the average case energy, and the Monte Carlo
  simulation. We used a $8\times 8\times 8$ system with both types of
  random interactions, bimodal ($J_{ij}=\pm 1$) and Gaussian
  distributed. }
\label{fig:PT-E3D}
\end{figure}

The Dual algorithm for the 3D EA model on the cubic lattice with no
redundant constraints converges for any temperature above $T\simeq
0.8$ and so we can use it to study the entire paramagnetic phase.  The
lack of convergence deep in the spin glass phase is to be expected.
Just as in the 2D case, the Dual algorithm (when converges) still
finds the same solution obtained by a Double Loop algorithm, and
again, it finds the solution nearly 100 times faster (see
Fig.~\ref{fig:timeDAIDual3D}).  Double Loop has the apparent advantage
of converging at any temperature even at very low ones. However, deep
in the spin glass phase, where the underlying paramagnetic
approximation is clearly inaccurate, we believe that an algorithm
(like the Dual one) that stops converging, is providing an important
warning that something wrong is probably happening. Such a warning
would be lacking by using a Double Loop algorithm.

In Figure \ref{fig:PT-Dual3D} the correlations predicted by the Dual
approximation, and those obtained by a Parallel Tempering Monte Carlo
simulation are compared. At high temperatures the correspondence is
quite good, but not as good as in 2D. However, it is important to
stress that the 3D EA model is much more difficult to simulate than
the 2D case: there is no exact method for computing the thermodynamics
(at variance to the 2D case) and Monte Carlo methods require huge
thermalization times, while the Dual algorithm runs in \emph{linear}
time with the system size.

In Figure~\ref{fig:PT-E3D} we show the estimates for the energy
obtained from the Monte Carlo and the Dual algorithm (both on a single
sample and on the average case).  The very strong agreement between
the Dual algorithm results on single samples and on the average case
is telling us that $U$-messages arriving at a given point on the
lattice are uncorrelated to a very large extent.  In other words, the
effect of short loops in the lattice is not manifestly present in
correlations between messages.  On the contrary, the comparison
between Dual algorithm results and Monte Carlo results is good only at
high temperatures, and it degrades when approaching the critical
temperature.  This discrepancy can be understood as due to a growing
correlation length in the EA model that diverges at the critical
temperature: our Dual approximation does not account for correlations
beyond the plaquette level and so it becomes inevitably poorer when
the correlation length diverges.  However, given the extremely fast
converging times of the Dual algorithm, it can be viewed as a very
effective algorithm for sampling the high temperature paramagnetic
phase and as a reasonable approximate algorithm when approaching the
critical point.

\section{Conclusions}
\label{Conclusions}

We have introduced a novel Dual algorithm to compute marginals
probabilities in the paramagnetic phase of frustrated spin models
(e.g.\ spin glasses) on finite dimensional lattices.  Inspired by the
fact that in a paramagnetic phase with no external field each variable
is unbiased (i.e.\ local magnetizations are null), the Dual algorithm
is derived by adding such paramagnetic constraints in the GBP
equations.  While BP (i.e.\ Bethe approximation) and GBP algorithms
have serious convergence problems at low temperatures even in the
paramagnetic phase, the Dual algorithm converges very fast in a much
wider range thanks to these constraints.  The Dual algorithm can also
be seen as BP on the dual lattice, where the interactions $J_{ij}$ act
as external fields on dual variables, thus improving convergence
properties of the message passing algorithm.

We have tested the Dual algorithm for the Edwards Anderson spin glass
model with bimodal and Gaussian couplings on 2D (square) and 3D
(cubic) lattices.  The results are very encouraging, showing
convergence in the whole paramagnetic phase (and even slightly in the
frozen phase for the 3D EA model) and comparing very well with exact
correlations measured in Monte Carlo simulations. A comparison with a
Double Loop algorithm (which is the state-of-the-art among general
purpose inference algorithms) shows that both algorithms found the
same result, but our Dual algorithm runs roughly 100 times faster. We
also tried to push the Dual approximation to the limit, and we used
the correlations inferred by the Dual algorithm to compute ground
states configurations in the 2D EA model by a freezing
procedure. Again, we showed that the ground states obtained in this
way compare very well with exact computations.

The success of our proposal clearly shows that as long as variables
are not long range correlated, the computation of correlations in a
generic spin model can be done in a very fast way by means of message
passing algorithms, based on mean-field like approximations. This kind
of inference algorithms do not provide in general an exact answer
(unless one uses it at very high temperatures or on locally tree-like
topologies), and so they can not be seen as substitutes for a Monte
Carlo (MC) sampling. However there are many situations where a fast
and approximate answer is required more that a slow and exact
answer. Let us just make a couple of examples of these situations. On
the one side, if one need to sample from very noisy data, an
approximated inference algorithm whose level of approximation is
smaller than data uncertainty is as valid as a perfect MC sampler. On
the other side, if one need to use the inferred correlations as input
for a second algorithm (as for the freezing algorithm in
Section~\ref{GroundState}) that will eventually modify/correct these
correlations, a fast and reasonably good inference is enough.

The promising results shown in the present work naturally ask for an
improvement in several directions.  For example, in the paramagnetic
phase of a model defined on a 3D lattice, our inference algorithm
could be improved by using the $2\times 2\times 2$ cube as the
elementary region, instead of the plaquette.  An even more important
improvement would be to extend the applicability range of the
algorithm to the low temperature phase: but this requires a rather non
trivial modification, since in low temperatures phase the assumption
of zero local magnetizations needs to be broken.

\begin{acknowledgments}
F. Ricci-Tersenghi acknowledges financial support by the Italian
Research Minister through the FIRB project RBFR086NN1 on ``Inference
and optimization in complex systems: from the thermodynamics of spin
glasses to message passing algorithms''.
\end{acknowledgments}

\end{document}